\documentstyle[12pt]{article}
\title{How to visualize a quantum transition \\of a single atom}
\author{
\large J\"urgen Audretsch\thanks{E-mail: 
Juergen.Audretsch@uni-konstanz.de} \and
Michael Mensky\thanks{Permanent address:
P.N.Lebedev Physical Institute, 117924 Moscow, Russia.\newline
E-mail: mensky@sci.lebedev.ru}
\and
Vladimir Namiot\thanks{Permanent address:
Institute of Nuclear  Physics, State Moscow University, 
117234  Moscow, Russia}
\\
\normalsize \it Fakult\"at f\"ur Physik der Universit\"at Konstanz\\
\normalsize \it Postfach 5560 M 674, D-78434 Konstanz, Germany}
\date{August 18, 1997}

\newcommand{\be}{\begin{equation}}
\newcommand{\ee}{\end{equation}}
\newcommand{\ba}{\begin{eqnarray}}
\newcommand{\ea}{\end{eqnarray}}

\newcommand{\eq}[1]{(\ref{#1})}
\newcommand{\Eq}{Eq.~\eq}

\newcommand{\ra}{\rangle}
\newcommand{\la}{\langle}

\newcommand{\al}{\alpha}
\newcommand{\const}{{\rm const}}
\newcommand{\Tlr}{T_{\rm lr}}
\newcommand{\dip}{{\rm dip}}
\newcommand{\r}{{\bf r}}
\newcommand{\Dd}{{\Delta d}}

\begin{document}
\maketitle

\begin{abstract}
The previously proposed visualization of Rabi oscillations of a single atom by a continuous fuzzy measurement of energy is specified for the case of a single transition between levels caused by a $\pi$-pulse of a driving field. An analysis in the framework of the restricted-path-integral approach (which reduces effectively to a Schr\"odinger equation with a complex Hamiltonian) shows  that the measurement gives a reliable information about the system evolution, but the probability of the transition becomes less than unity. In addition an experimental setup is proposed for continuous monitoring the state of an atom by observation of electrons scattered by it. It is shown how this setup realizes a continuous fuzzy measurement of the atom energy. 
\end{abstract}

\section{Introduction}

The problem of quantum measurements and related questions of decoherence, wave packet reduction etc. have always found great interest. In the last years continuous measurements were considered intensively 
\cite{ContinMeas}-\cite{PrevPaper}. 
In this connection the so-called quantum Zeno effect has been predicted and then confirmed experimentally \cite{Zeno}. This effect shows how a continuous measurement prevents transitions between discrete spectrum states. It thus demonstrates in the most evident way that a continuous measurement may strongly influence the measured system.  

In real measurements this influence on the measured system does not lead to absolute freezing its evolution but only strongly modifies it \cite{Milburn}. In the specific setup of a three-level system a random telegraph-type signal may be obtained from the measured system as a consequence of its ``shelving" \cite{GisinKnightPercivalThompson}. In all these cases the evolution of the discrete-level measured system is radically modified by the measurement as a result of strong back influence of the measuring device onto the measured system. The reason for this strong influence is that the considered continuous measurement was in fact a series of often short measurements which were so strong that they projected the measured system on one of its discrete eigenstates. In \cite{JacobsKnight} such a procedure is called a ``continuous projection measurement". 

In this context it is interesting and important to study a different class of quantum measurements and to ask the complementary question: Is it possible to continuously measure an individual quantum system with a not too strong influence on it so that the behavior of its state does not radically differ from what it would be if no measurement is performed. It is then to be expected that the obtained continuous measurement readout reflects the motion of the state thus making it visible. For this aim the measurement must be weak enough. Correspondingly it is unavoidable that it must have  a not too high resolution, since the better the resolution of a measurement is, the stronger is its influence on the measured system. Because of this property such a weak measurement will also be called fuzzy. If now the measurement is presented as a series of short measurements, each of them must be fuzzy enough not to project on a single state. 

Finite-resolution continuous measurements in general as well as their influence on measured systems were in great detail investigated in \cite{RPI} in the framework of the phenomenological restricted-path-integral (RPI) approach. The continuous fuzzy measurement of an observable with discrete spectrum has been investigated in \cite{Peres-book} in the context of the ensemble approach. It was shown that averaging over the ensemble of many readouts gives an information about the behavior of the state of the system (for example about Rabi oscillations). However this procedure is not applicable to an individual system like a single atom which is continuously measured. 

The RPI approach on the other hand is. It  was first applied to a discrete-spectrum observable (energy of a two-level system) in \cite{Onofrio-energy}. It was shown that the quantum Zeno effect arises if the resolution of the measurement is good enough (in comparison with the level difference). However the analysis given in \cite{Onofrio-energy} was not complete because only the special case of constant measurement readouts coinciding with energy levels ($E(t)\equiv \const = E_n$) was considered. 

The detailed analysis given in \cite{PrevPaper} showed that in the case of the measurement with an intermediate resolution (not too low to give no information and not too high to lead to the Zeno effect) transitions between levels (Rabi oscillations) maintain though are modified, and the measurement readout $E(t)$ is correlated with these oscillations. Thus, transitions between levels may be continuously monitored with the help of a fuzzy continuous measurement. Of course, the error of this monitoring is comparatively large and principally cannot be made small. 

In the present paper we shall continue exploring this possibility in a simple case of a so-called $\pi$-pulse of the driving field bringing the system, in absence of any measurement, from one level to another. We shall discuss a continuous fuzzy measurement of energy in such a system. It will be shown that the measurement readout visualizes the quantum transition but on the price that the transition becomes less probable. We shall first shortly discuss the problem in the framework of the RPI approach to show what results of the measurement may be expected in this case. Then a possible experimental realization of the measurement of this type, namely, monitoring the energy of an atom by scattering electrons on it, will be described. 

\section{Predictions made by the RPI method}\label{sect-meas-trans}

We consider a multilevel system with Hamiltonian $H_0$ and corresponding energy levels $E_n$. It is influenced by an external  driving field with potential $V=V(t)$ leading to transitions if no measurement is applied. Then the continuous measurement of the energy $H_0$ of this system may be described according to \cite{PrevPaper} by the effective Schr\"odinger equation 
\be
\frac{\partial}{\partial t} |\psi_t\rangle
  = \left(-\frac{i}{\hbar} H
  -\kappa \,\Big( H_0 - E(t)\Big) ^2\right)\, |\psi_t\rangle
\label{efSchroedEnergy}\ee
with $H=H_0+V$. For a given measurement readout $[E] = \{ E(t)|T_1\le t \le T_2\}$ the probability density for the realization of this readout is given by $P[E]=\la\psi_{T_2}|\psi_{T_2}\ra$ where $|\psi_{T_2}\ra$ is the solution $|\psi_t\ra$ of \Eq{efSchroedEnergy} taken at the end of the observation interval $[T_1,T_2]$. Note that because of the damping term the norm of $|\psi_{T_2}\ra$ decreases with time. The inverse of the parameter $\kappa > 0$ may serve as a measure of fuzziness. No measurement corresponds to $\kappa = 0$. Normalization of $|\psi_{T_2}\ra$ leads to the state vector at $t =T_2$. 

Expanding $|\psi_t\ra$ according to $|\psi_t\ra = \sum C_n(t)|\varphi_n(t)\ra$ in the orthonormal basis $|\varphi_n(t)\rangle = e^{-iE_n\, t/\hbar}|n\rangle$ of time-dependent eigenstates of $H_0$, we obtain $P[E]=\sum_n|C_n(T_2)|^2$. For the case of a 2-level system and a resonant driving field with frequency $\omega$ ($\Delta E = E_2 - E_1 = \hbar\omega$) \Eq{efSchroedEnergy} reduces to 
\begin{eqnarray}
\dot C_1 &=&  -i v(t) C_2 - \kappa (E_1-E(t))^2\, C_1,
\nonumber\\
\dot C_2 &=&  -i v(t) C_1 - \kappa (E_2-E(t))^2\, C_2
\label{2levelEq}\end{eqnarray}
with $\langle\varphi_1|V\varphi_2\rangle=\langle\varphi_2|V\varphi_1\rangle^{*}=\hbar v(t)$. 

For $\kappa = 0$ and $v(t) = v_0 = \const$ the system undergoes Rabi oscillations between the eigenstates with the period $T_R=\pi/v_0$. We consider a $\pi$-pulse in the interval $[0,T]$ where $T_1\le 0\le T\le T_2$,  so that $T=T_R/2$ and $v(t) = v_0$ in the interval  $[0,T]$, $v(t) = 0$ outside this interval. 

If the system is on level 1 at the initial time moment ($C_1(T_1)=1$, $C_2(T_1)=0$) and no measurement is performed ($\kappa = 0$ so that the wave function is normalized at all times), then the system is subject to the level transition  during the interval $[0,T]$. This means that the probability $|C_2(t)|^2$ to be at level 2 gradually increases during this interval and achieves unity at the end of it. 

A quantity by which the fuzziness of the measurement can be described quantitatively and which may replace $\kappa$ in physical discussions is the level resolution time $\Tlr = 1/\kappa\Delta E^2$ (comp. \cite{PrevPaper}). If a continuous fuzzy measurement lasts longer than $\Tlr$, it distinguishes (resolves) between the levels. Small $\Tlr$ represents quick level resolution because of small fuzziness and therefore strong Zeno-type influence of the measurement. The larger $\Tlr$ is, the weaker is the influence of measurement on the atom.

It was shown in \cite{PrevPaper} that with $T_R$ and $4\pi \Tlr$ being of the same order there is a regime where correlations between oscillations of the state vector and the energy readout $[E]$ are to be expected. This regime lies between the Zeno and  Rabi regimes. In extending \cite{PrevPaper} we present now some results of a numerical analysis of \Eq{2levelEq} for this regime of measurement. Details will be published elsewhere. 

Random curves $E(t)$ were generated, for each of them \Eq{2levelEq} was solved, the norm of the solution at the final time $|C_1(T_2)|^2+|C_2(T_2)|^2$ was found and interpreted as the probability density $P[E]$ that the curve $E(t)$ occur as the measurement readout. The corresponding normalized coefficients $c_n=C_n/\sqrt{|C_1|^2+|C_2|^2}$ present the behavior of the system for the case that the measurement gives the readout $E(t)$. 

For the further analysis all curves $E(t)$ were smoothed \label{smoothing} with the time scale of the order of $T$ to eliminate insignificant fast oscillations but conserve the information about the transition. Then they were separated into 4 classes $E_{11}$, $E_{12}$, $E_{21}$, $E_{22}$ according to the location of initial $E(T_1)$ and final $E(T_2)$ points of the curve. For example, if the curve begins below the middle line $\overline{E} = (E_1 + E_2)/2$ and ends higher than this, it is included in the class$E_{12}$ and interpreted as the readout describing a transition from the level 1 onto level 2. Probabilities of these classes were found by summation of probability densities $P[E]$ by the Monte Carlo method. 

In the case of a $\pi$-pulse one could naively expect that only curves of the class $E_{12}$ may arise as measurement readouts (as is the case without the measurement). The calculation gives different results. For the regime of measurement intermediate between Zeno and Rabi the results are following. The curves of the classes $E_{21}$ and $E_{22}$ arise with small probabilities.  However the probabilities of the classes $E_{11}$, $E_{12}$ are comparable. Therefore, influence of the measurement sometimes may prevent the transition.  

\begin{figure}
\let\picnaturalsize=N
\def\picsize{8cm}
\def\picfilename{scat1.eps}
\ifx\nopictures Y\else{\ifx\epsfloaded Y\else\input epsf \fi
\let\epsfloaded=Y
\centerline{\ifx\picnaturalsize N\epsfxsize \picsize\fi \epsfbox{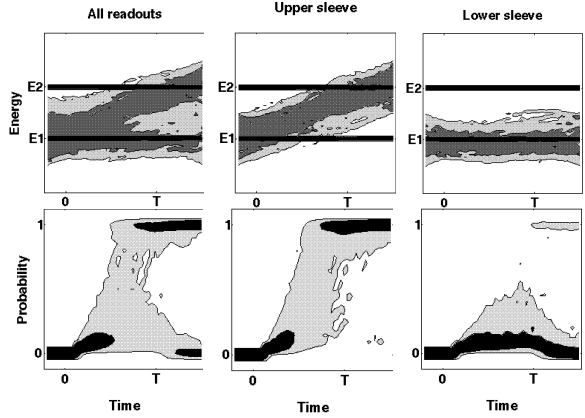}}}\fi
\caption{\rm Density plots of measurement readouts $E(t)$ and corresponding curves $|c_2(t)|^2$ giving the probability to be at level 2. In the left pair of diagrams all measurement readouts are drawn while the middle and right ones represent the readouts of the classes $E_{12}$ (transition from level 1 to level 2) and $E_{11}$ (staying at level 1) correspondingly.}
\label{figRPIdens}\end{figure}

Fig.~\ref{figRPIdens} gives an idea of what the result of the measurement will be and how it will reflect the behavior of the system in an intermediate measurement regime ($4\pi \Tlr / T_R = 5/3$).
Upper diagrams present possible readouts $E(t)$ and the corresponding curves $P_2(t)=|c_2(t)|^2$ characterizing the behavior of the system state are given below. The left pair of diagrams shows all possible readouts $E(t)$, while the middle and right pairs show the readouts of the classes $E_{12}$ and $E_{11}$ correspondingly. The curves are presented by density plots  taking into account probability densities $P[E]$ (being the same for the curve $E(t)$ and the corresponding $P_2(t)$). 

It is seen from Fig.~\ref{figRPIdens}  that the information given by the measurement readout $E(t)$ is reliable. This means that a readout $E(t)$ and the corresponding curve $P_2(t)$ characterizing the behavior of the measured system with high probability correspond to each other. For example, if $E(t)$ belongs to the class $E_{12}$, then  the behavior of $P_2(t)$ is characteristic for the transition from level 1 to level 2. If $E(t)$ belongs to the class $E_{11}$, then  $P_2(t)$ also in most cases describes the system which finally stays at level 1 or close to it.  

Thus, the qualitative conclusion from the numerical analysis is the following: the visualization of an externally driven quantum transition by a continuous measurement of energy of the appropriate fuzziness is possible. But as price for this information the transition itself becomes uncertain: it may occur or not occur with probabilities of the same order. 

The probability of the transition increases with the measurement becoming more fuzzy (larger $4\pi \Tlr$). However, the information becomes then less reliable, the noise increases. When the measurement becomes too fuzzy ($4\pi \Tlr\gg T_R$), the Rabi regime arises: the measurement does not influence the system but it also gives no information. 

Vice versa, the transition becomes less probable for a more strong measurement (smaller $4\pi \Tlr$). By too strong measurement ($4\pi \Tlr\ll T_R$) the transition is prevented completely and the Zeno regime of the measurement arises. 

Let us now turn to a real experimental setup where these theoretical predictions can be observed. 

\section{The experimental setup}

The basic idea for the realization of a fuzzy measurement of energy proposed below is the following: A single 2-level atom with energy levels $E_1$ and $E_2$ changes its state 
\be
\psi = \left( 
\begin{array}{c}
c_1(t)\\
c_2(t)
\end{array}
 \right)
\label{WaveFunct}\ee
(normalized so that $|c_1(t)|^2+|c_2(t)|^2=1$) under the influence of a resonant laser field. We want to monitor some characteristic of the state of the atom that may be interpreted as monitoring of its mean energy given by 
\be
E(t) = E_{1}|c_{1}(t)|^{2} + E_2|c_{2}(t)|^{2}
\label{En-C1C2}\ee

We assume that the polarizabilities $\al_1$ and $\al_2$ of the atom on the two levels\footnote{In the framework of the two-level approximation, the polarizabilities may be introduced only phenomenologically. A correct definition of them would require to include the other levels of the atom.} are different. The atom is placed in an electric field ${\cal E}_0$ which induces a dipole moment. Since the polarization of the atom in the process of its transition between the energy eigenstates changes, the dipole moment changes correspondingly. 

If we now scatter a sequence of single electrons by this atom, the corresponding cross section will depend on $(\al_2 - \al_1) {\cal E}_0 (|c_2(t)|^2-|c_1(t)|^2)$ and can be determined as a function of time. This results according to \eq{En-C1C2} in an energy readout $E(t)$. As we will show below, there are means of controlling the weakness of the back influence of the electrons on the atom, so that fuzzy measurements are possible. 

\begin{figure}
\let\picnaturalsize=N
\def\picsize{8.5cm}
\def\picfilename{scat2.eps}
\ifx\nopictures Y\else{\ifx\epsfloaded Y\else\input epsf \fi
\let\epsfloaded=Y
\centerline{\ifx\picnaturalsize N\epsfxsize \picsize\fi \epsfbox{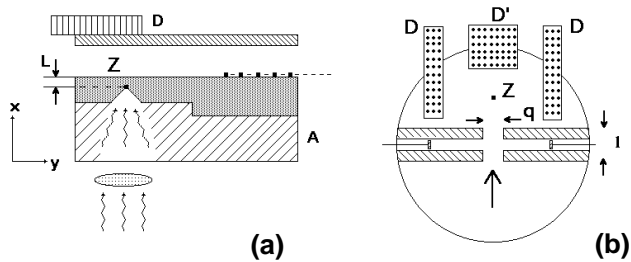}}}\fi
\caption{\rm The experimental setup from the side (a) and from above (b): A single atom $Z$ is implanted on a substrate $A$. Laser radiation is focussed on it. Electrons swimming on the surface of liquid helium (black) are collimated and scattered one by one by the atom. They are registered by the microchannel plates $D$ and $D'$ as being deflected or not.}
\label{fig-exper}
\end{figure}

In the experimental setup which we propose (see Fig.~\ref{fig-exper})
an atom $Z$ is implanted on a substrate $A$. A constant electric field ${\cal E}_0$ directed along the $x$-axis is applied to induce a  dipole moment. This field is created by a molecule with a fixed dipole moment $d_1$ (also directed along $x$) located near the atom $Z$. The resonant radiation of frequency $\omega$ is focused on the atom. It causes a dipole moment consisting of two contributions: a comparatively slowly varying one to which we referred above and quickly oscillating one. The latter is proportional to $|c_1^*(t)c_2(t)|d'$ where $d'$ is the dipole moment of the transition between the first and second levels of the atom. The characteristic frequency of oscillations is high ($\omega\sim 10^{15}$sec$^{-1}$)  and therefore its contribution to the scattering cross section of the electrons is negligible. 

Substrate and the atom are covered by a thin layer of liquid helium. The possibility to use liquid helium for the exploration of single microobjects has been considered earlier in \cite{NamiotHelium}. Free electrons swim on the surface of liquid helium \cite{ElectrLiqHel,4} being pressed to this surface by a constant electric field produced by a plate above the helium. Electrons are scattered by the electric field of the dipole moment of the atom. Moving along the surface of the liquid helium after scattering,  the electrons come to the region under microchannel plates capable to register single electrons (the potential between the plates and the substrate creates an electric field  turning the electron off the surface and accelerating them toward the plates). Different problems regarding the realization of the experiment will be addressed below. 

\section{Analysis of the experiment}

The electrons swimming on the surface of the liquid helium are scattered by a potential $V_{\dip}(\r)$ which is caused by the fixed external dipole moment $d_1$ and the induced atomic dipole moment:
\be
V_{\dip}(\r) = \frac{eL}{(L^2 + \r^2)^{3/2}}
\left[ 
d_1 + 
\left(
\begin{array}{cc}
\alpha_1 & 0 \\
0             & \alpha_2          
\end{array}
\right) {\cal E}_0
\right]
\label{Interact}\ee
where $L$ denotes the distance between the atom and the surface of the liquid helium (comp. Fig.~\ref{fig-exper}a), $\r$ is the two-dimensional vector describing the position of the electron on the surface and $e$ is the electron charge. 

The atom and the electron passing by are treated as a composed system which is described in the corresponding product space. In a scattering process the state of the electron as well as the state of the atom is changed. Let us address the electron first. In the Born approximation \cite{LandauLifsh}  in two dimensions the differential cross section for the electron scattering (which has the dimension of length) turns out to be 
\be
\frac{d\sigma}{d\theta} = \beta(t)\exp \left[  -\gamma \sin \left(\frac{\theta}{2}\right) \right]
\label{dif-section}\ee
where $\beta(t) = 2\pi m^{3/2}e^2 d^2(t) / \hbar^3(2E_e)^{1/2}$, $\gamma = 4L(2E_e m)^{1/2}/\hbar$ ($m$ is the electron mass, $E_e$ its energy and $\hbar$ Planck constant). The time dependent state of the atom enters \Eq {dif-section} via
\be
d^2(t) = d_0^2 + \Dd^2 + 
2d_0\Dd (|c_2(t)|^2 - |c_1(t)|^2)
\label{dipole-zero}\ee
where
\be
d_0 = d_1 + \frac 12 (\alpha_2 + \alpha_1) {\cal E}_0 , \quad
\Dd =  \frac 12 (\alpha_2 - \alpha_1) {\cal E}_0
\label{dipole-not}\ee

The total cross section corresponding to  \Eq {dif-section} is 
\be
\sigma(t) = 2\pi\beta(t)[I_0(\gamma) - L_0(\gamma)]
\label{compl-section}\ee
where $I_0(\gamma)$ is the Bessel function with an imaginary argument and $L_0(\gamma)$ the modified Struwe function \cite{Functions}. However a simpler expression may be taken to obtain estimates: 
\be
\sigma(t) = \frac{4\beta(t)}{\gamma} \left[  1 - \exp\left( -\frac{\pi\gamma}{2}\right) \right]
\label{compl-section-simple}\ee
We have $\sigma\sim 10^{-7}$m for $d(t)\sim 10^{-29}$Coulomb m, $E_e \sim 10^{-4}$eV. 

To have no additional complications let us restrict to the case where $\gamma$ is sufficiently small ($\gamma\ll 1$) so that  $d\sigma/d\theta$ does not depend on $\theta$. To avoid the dispersion of $d\sigma /d\theta$ due to the dispersion of energies of the incoming electrons, the electrons with energies strongly different from some fixed energy $E'_e$ should not participate in the scattering. This may be provided by different technical ways which we will not discuss in the present paper. We shall only introduce a coefficient $s$ indicating the number of times the primary flux of electrons is attenuated after such separation. 

The total time dependent cross section for the electron scattering has then the form $\sigma(t)=\sigma_0 + \delta\sigma(t)$ where only the second term depends on the momentary state of the atom:
\be
\sigma_0 =  \frac{4\pi^{2}m^{3/2}e^{2}}{\hbar^{3}(2E'_e)^{1/2} } 
(d_0^2 + \Dd^2), \quad 
\delta\sigma(t) = \chi(|c_2(t)|^2 - |c_1(t)|^2)
\label{sigma-deltasig}\ee
with
\be
\chi =
\sigma_{0}
\frac{2d_0\Dd }{ d_0^2 + \Dd ^2}
\label{chi}\ee

We add some remarks concerning the experimental realization. The applicability of the Born approximation requires validness of the following condition \cite{LandauLifsh}:
\be
\frac{ed(t)}{L^2} \ll \frac{\hbar^2}{mL^2} \left( \frac{2mE_e}{\hbar^2} \right) \; L
\label{Born-condition}\ee
Both parts of Eq.~(\ref{Born-condition}) turn out to be of the same order for realistic parameters ($L\sim 10^{-8} $m, $E_e\sim 10^{-4}$eV, $d(t)\sim 10^{-29}$Coulomb m). Eq.~(\ref{dif-section}) may nevertheless be used at least for rough estimates. 

Let us estimate the flux $F(t)$ of the scattered electrons which previously passed through the collimator consisting of two slits of the width $q$ with the distance $l$ between them: 
\be
F(t) \sim \frac ql \, s  \, \sigma(t) n_e \left(  \frac{2E'_e}{m} \right)^{1/2} = g\sigma(t)
\label{flux}\ee
Here $g$ is the flux of the incoming electrons. It is expressed in the following parameters: $s$ is a factor accounting for attenuation of the flux after selection of their energy (see above), $n_e$ is the number of electrons per a unit of area of the surface of the liquid helium. To be able to distinguish the scattered electrons from those not scattered, it is necessary that the ratio $q/l$ satisfies the condition $q/l \ll \gamma^{-1}$. For the parameters indicated above, we can take $q/l \sim 0.1$. The estimate for $s$ is $s\sim\delta E_e/E'_e$ where $\delta E_e$ is the dispersion of energies of incoming electrons. Taking $\delta E_e\sim 0.1E'_e$ we have $s\sim 0.1$. Then, for $n_e\sim 10^{13}$m$^{-2}$ (an instability arises for larger $n_e$ because electrons ``drown" in the liquid helium \cite{4}), $E'_e\sim 10^{-4}$eV and $s\sim q/l \sim 0.1$ we have $F(t) \sim (10^{6}\div10^{7})$sec$^{-1}$. If we investigate some transition of the atom between its two energy eigenstates, it is evident that sufficiently large number of scatterings has to occur during the time $T$ of the  transition. In other words, the inequality $F(t)T\gg 1$ must be fulfilled. On the other hand, $T$ must be less than the time of spontaneous radiance of the atom $Z$. Based on the above estimate of $F(t)$, the time of spontaneous radiance must be significantly larger than ($10^{-6}\div 10^{-7}$)sec. This imposes restrictions on the choice of possible candidates for the atom $Z$. 

We assume that the ingoing electrons are coming one by one and are either deflected or not deflected by the atom. Before such an event happens at the time $t$ the atom is in the state $\psi(t)$ of \Eq{WaveFunct}. After the event the total system is in an entangled state for which the deflected electron state and the non-deflected electron state are combined with the respective atom states after the electron has passed by. The electrons are then registered as being deflected or not by the corresponding plates $D$ and $D'$ (see Fig.~\ref{fig-exper}b). This reduces the state of the total system and the state of the atom is changed this way. If a deflection of the electron happens, the influence of the interaction potential $V(\r)$ of \Eq{Interact} results in the modified normalized atom state $\psi'(t)$ with 
\be
\left( 
\begin{array}{c}
c'_1\\
c'_2
\end{array}
 \right) = 
\left(
\begin{array}{cc}
a_1(t) & 0 \\
0             & a_2(t)          
\end{array}
\right) 
\left( 
\begin{array}{c}
c_1\\
c_2
\end{array}
 \right) 
\label{wave-func}\ee
whereby 
\be
a_1(t) = \frac{d_0 - \Dd }{d(t)}, \quad
a_2(t) = \frac{d_0 + \Dd }{d(t)}
\label{v12scattered}\ee

If on the other hand the deflection did not happen (i.e. the electron was registered by the microchannel plate $D'$), the expression for $a_1(t)$, $a_2(t)$ has another form: 
\ba
a_1(t) &=& 
\left[ 
1 - \frac{\sigma(t)(d_0 - \Dd )^2}{2qd^2(t)}
\right]
\left/
\sqrt{ 
1 - \frac{\sigma(t)}{q} 
}
 \right.
\nonumber\\
a_2(t) &=& 
\left[ 
1 - \frac{\sigma(t)(d_0 + \Dd )^2}{2qd^2(t)}
\right]
\left/
\sqrt{ 
1 - \frac{\sigma(t)}{q} 
}
 \right.
\label{v12non-scattered}\ea

To sum up: The observation of the scattered electrons at the time $t$ supplies some information about $\sigma(t)$ (see Sect.~\ref{sect-fuzziness} for details). By this an information about the state $\psi(t)$ of the atom (namely about $|c_2(t)|^2 - |c_1(t)|^2$) is obtained by which the energy $E(t)$ of the atom may be worked out according to 
\be
E(t) = \overline{E} + \frac{\Delta E}{2\chi}\delta\sigma(t)
= \overline{E} + \frac{\Delta E}{2\chi}(\sigma(t) - \sigma_0)
\label{En-sigma}\ee
where $\overline{E} = (E_1 + E_2)/2$ and $\Delta E = E_2 - E_1$. This formula enables us to compare the theoretically calculated behavior of the function $E(t)$ with the function $\sigma(t)$ determined from the experimental scattering results. Simultaneously the state of the atom is changed. It is important to note that according to Eqs.~\eq{wave-func} and \eq{v12scattered} this must not result in a projection onto one of the energy eigenstates. Rather the atomic state may be only weakly influenced in one scattering event if $|\Dd| \ll d_0$. This represents a condition for the measurement of $E(t)$ to be fuzzy. We return to this below. Let us first proceed to the question how to measure $\sigma(t)$. 

\section{Degree of fuzziness}\label{sect-fuzziness}

We have to address the problem how to estimate $\sigma(t)$ from the number of scatterings.  We make this estimate on a series consisting of $N$ events when a single electron passes the atom. Let $N_1(t)$ be the corresponding number of events in the series resulting in the deflection. 

Before scattering, the electrons have to pass the slit of the collimator with the opening $q$ (see Fig.~\ref{fig-exper}b).  The ratio $\sigma(t)/q$ is equal to the probability $p$ that a single electron will be scattered by the atom. This probability may be estimated from the data on $N$ electrons by the ratio $N_1/N$ where $N_1$ is the number of those electrons which were deflected. We then have approximately 
\be
\frac{\sigma(t)}{q} \approx \frac{N_1(t)}{N}
\label{sigma-n}\ee
or with \eq{En-sigma} 
\be
\frac{E(t) - \overline{E}}{\Delta E} \approx    \frac{q}{2\chi}\left(\frac{N_1}{N} - \frac{\sigma_0}{q}\right)
\label{En-N}\ee
This completes the measurement of $E(t)$ of \Eq{En-C1C2}. 

In measuring $\sigma(t)$ in fact $|c_2|^2$ will be measured in the experiment. This corresponds indeed to the intended visualization of the quantum transition. But with the setup described above one is indeed capable of more, namely of testing experimentally evidence for the theoretical predictions for the continuous fuzzy measurement of energy as they are illustrated by Fig.~\ref{figRPIdens}. The reason for this is that a detailed analysis of the setup (which will be presented elsewhere) shows that this setup represents a continuous fuzzy measurement of energy in the sense of Sect.~\ref{sect-meas-trans} provided that the difference of the dipole moments attributed to the two energy eigenstates is small enough, 
\be 
|\Dd | \ll d_0
\label{small-dipole-dif}\ee
The level resolution time (a measure of fuzziness) turns out to be then
\be
\Tlr = \frac{1}{4g\sigma_0}
\left(
\frac{d_0}{\Dd }
\right)^2
\left( 1 - \frac{\sigma_0}{q} \right)
\label{Tlr-sigma}\ee

As discussed already in connection with Eqs.~\eq{wave-func} and \eq{v12scattered}, the second factor in \Eq{Tlr-sigma} represents the weakness of the influence on the atom caused by the deflection of one single electron. In fact we are concerned with a succession of many scattering events. If they happen very quickly, $g\sigma_0$ is large. Correspondingly the influence of the measurement will be strong and $\Tlr$ small. This explains the first factor in \Eq{Tlr-sigma}. 

\section{Conclusion}

The previously predicted possibility to visualize the time dependence of a quantum transition of a single atom with the help of a fuzzy continuous measurement of energy is demonstrated in detail for the case of a $\pi$-pulse of a driving field. The realization of the corresponding measurement is discussed. The RPI analysis shows that the visualization is possible but at the price of additional quantum uncertainties. Namely, even in the case of a $\pi$-pulse, when in the absence of the measurement the transition occurs with certainty, it may with comparable probabilities occur or not in the presence of the continuous fuzzy measurement. In both cases the information given by the energy readout is reliable i.e. it corresponds with high probability to the evolution of the system state. 

As a scheme of realization, scattering electrons by a fixed two-level atom  is considered. The electrons probe the state of the atom. The fuzziness (weakness) of the measurement is represented by the fact that each scattering event changes the internal state of the atom only insignificantly. This is radically different from the often considered scheme of repeated measurements in which the atom is projected on one of the levels after each measurement. 

\vskip 0.5cm
\centerline{\bf ACKNOWLEDGEMENT}

\noindent
This work was supported in part by the Deutsche Forschungsgemeinschaft. M.M. is thankful to P.L.Knight who presented a manuscript of his paper before publication.


\begin{thebibliography}{99}
\bibitem{ContinMeas} 
H.D.Zeh, 
Found. Phys. {\bf }1, 69 (1970); {\bf 3}, 109 (1973); 
E. B. Davies, {\em Quantum Theory of 
Open Systems}, Academic Press: London, New York, San Francisco, 
1976; 
M. D. Srinivas, {\em J.Math. Phys.} {\bf 18}, 2138 (1977);
A. Peres, Continuous monitoring of quantum 
systems, in {\em Information Complexity and Control in Quantum 
Physics}, ed. by A. Blaquiere, S. Diner, and G. Lochak, Springer, 
Wien, 1987, pp. 235;
D. F. Walls and G. J. Milburn, Phys. Rev. 
{\bf A 31}, 2403 (1985); 
E. Joos, and H. D. Zeh, Z.Phys. {\bf B 59}, 223 (1985);
L. Diosi, Phys. Lett. {\bf A 129}, 419 (1988);
A. Konetchnyi, M. B. Mensky and V. Namiot, 
Phys. Lett. {\bf A 177}, 283 (1993);
P.Goetsch and R .Graham, Phys. Rev. {\bf A~50}, 5242 (1994); 
T.Steimle and G.Alber, Phys. Rev. {\bf A~53}, 1982 (1996).

\bibitem{RPI} 
M. B. Mensky, Phys. Rev. {\bf D 20}, 384 (1979); 
Sov. Phys.-JETP {\bf 50}, 667 (1979);
M.B.Mensky, {\em Continuous Quantum Measurements
and Path Integrals}, IOP Publishing, Bristol and Philadelphia, 1993.

\bibitem{Zeno} B.Misra and E.C.G.Sudarshan, J. Math. Phys. {\bf 18}, 756 (1977); 
C.B.Chiu, E.C.G.Sudarshan and B.Misra,
Phys. Rev. {\bf D~16}, 520 (1977); 
A.Peres, Amer. J. Phys. {\bf 48}, 931 (1980); 
F.Ya.Khalili, Vestnik Mosk. Universiteta,
ser. 3, no.5, p.13 (1988); 
W.M.Itano, D.J.Heinzen, J.J.Bollinger and
D.J.Wineland, Phys. Rev. {\bf A~41}, 2295 (1990); 
A.Beige and G.C.Hegerfeldt, 
Phys. Rev. {\bf A~53}, 53 (1996). 

\bibitem{Milburn} 
M.J.Gagen and G.J.Milburn, 
Phys. Rev. A47, 1467, 1993. 

\bibitem{GisinKnightPercivalThompson} 
N.Gisin, P.L.Knight, I.C.Percival and R.C.Thompson, 
J. Modern Optics {\bf 40}, 1663 (1993). 

\bibitem{JacobsKnight}
K.Jacobs and P.L.Knight, 
Linear quantum stochastic equations: Applications to continuous projection measurements, 
Phys. Rev.~{\bf A}, to be published. 

\bibitem{Peres-book} A. Peres, {\em Quantum Theory: Concepts and 
Methods}, Kluwer Academic Publishers, Dordrecht, Boston \& 
London, 1993.

\bibitem{Onofrio-energy} 
R. Onofrio, C. Presilla, and U. Tambini, Phys. 
Lett. {\bf A 183}, 135 (1993); 
U. Tambini, C. Presilla, R. Onofrio, Phys. 
Rev. {\bf A 51}, 967 (1995).

\bibitem{PrevPaper} J.Audretsch and M.B.Mensky, Phys. Rev. {\bf A~56},  44 (1997).

\bibitem{Lippert} Ernst Lippert and J.D. Macomber, {\em Dynamics during spectroscopic transitions. Basic concepts}, Berlin and Heidelberg, Springer
(1995). 

\bibitem{NamiotHelium}
V.A.Namiot, Phys. Lett. {\bf A~230}, 126 (1997). 

\bibitem{ElectrLiqHel} 
M.W.Cole, H.Cohen, Phys. Rev. Lett. 23, 1238 (1969); 
V.B.Shikin and Yu.P.Monarkha, "Fizika Nizkikh Temperatur" {\bf 1}, 958(1975); 
V.S.Edelman, Zh. Exp. Teor. Fiz.  77, 673 (1979). 

\bibitem{4} A.P.Volodin, M.S.Khaikin, and V.S.Edelman,  Pis'ma Zh. Exp. Teor. Fiz.   26, 707 (1977). 

\bibitem{LandauLifsh} L.D.Landau and E.M.Lifshits, {\em Quantum Mechanics. Nonrelativistic theory}, Oxford, Pergamon Press, 1987. 

\bibitem{Functions} M.Abramowitz, {\em Handbook of Mathematical Functions}, Dover Publ., New York, 1972. 

\end{thebibliography}
\end{document}